# Electron configuration of the [FeO]$^{2+}$ group in the H-abstraction from methane: oxyl Fe$^{III}$-O$^{\bullet}$ versus ferryl Fe$^{IV}$=O

Igor L. Zilberberg[1]

*Boreskov Institute of Catalysis, Novosibirsk 630090, Russian Federation*

This account presents author's opinion on the mechanism of the H-abstraction from methane by the [FeO]$^{2+}$ group. In the course of reaction with hydrogen, the Fe-O bond in the ferryl Fe$^{IV}$=O configuration becomes elongated causing transfer of the α-spin electron from one of doubly occupied π-bonding orbitals leaving behind single β-spin electron on oxygen. This oxygen in so-formed oxyl Fe$^{III}$-O$^{\bullet}$ configuration of the [FeO]$^{2+}$ moiety then easily accept the α-spin hydrogen atom from methane in the same way as the radical-localized oxygen does. This mechanism is compared with the scheme in which the hydrogen is accepted by low-lying unoccupied σ*(Fe-O) orbital in the ferryl configuration.

## Introduction

The [FeO]$^{2+}$ group is known to be key intermediate in the oxidation catalysis by the non-heme iron enzymes. [1],[2] The same group formed by dissociation of N$_2$O on ferrous site in zeolite Fe-ZSM-5, called "α-oxygen center", revealed extraordinary activity in the methane-to-methanol oxidation. [3] The electron configuration of this center is suggested by these authors as Fe$^{III}$-O$^{\bullet-}$ (recently they switched to designation of this state as Fe$^{III}$-O$^{\bullet}$)[4] thus assigning the oxyl radical state to the terminal oxygen on base of the reactivity data for this species. The lack of direct structural data for α-oxygen creates some uncertainty in understanding the mechanism of the C-H bond activation on it. In particular, for the Fe$^{III}$-O$^{\bullet}$ model, there have been unknown the factors stabilizing the radicaloid oxygen, i.e. prevent the oxygen in this species from taking a second electron from ferric iron to return to ferryl state Fe$^{IV}$=O. The existing quantum-chemical studies of α-oxygen seem to be still far from consistent explanation of extraordinary reactivity of this oxidant.

Consensus is not achieved yet even on nature of the ground state for the [FeO]$^{2+}$ species: whether it is oxyl Fe$^{III}$-O$^{\bullet}$ as suggested by Panov with coworkers, or ferryl Fe$^{IV}$=O. Only the latter configuration is considered as ground-state one for the [FeO]$^{2+}$ species in the coordination chemistry of dianionic oxo ligands [5], Fenton's chemistry [6], iron-oxygen active species within

---
[1] **E-mail:** I.L.Zilberberg@catatalysis.ru

CuFe-ZSM-5 zeolite mediated catalytic oxidation of methane to methanol with $H_2O_2$ under benign conditions [7], and numerous mono-iron complexes in biomimetic chemistry. [1],[2]

The assignment for α-oxygen to the $Fe^{III}$-$O^{•}$ configuration in the ground state of the system was supported by means of the resonant inelastic X-ray scattering method which provided the evidence for the pure $3d^5$ configuration of the iron center in the $[FeO]^{2+}$ group.[8] Quite recently, Solomon with co-authors made a contradicting assignment of α-oxygen to the $Fe^{IV}$=O electron structure obtained on the base of variable-temperature variable-field magnetic circular dichroism data in conjunction with the CASP2/B3LYP calculations.[9]

Despite the above mentioned structural uncertainty for the α-oxygen center, the designed cluster models (developed by Zhidomirov with coworkers [10]) allowed one to make some suggestions. One of such model in which the $[FeO]^{2+}$ group is placed in the cavity of zeolite six-membered ring with two Al atoms appeared quite useful.[10] For this model, the ferryl-type ground state was identified by Baerends with coworkers.[11]

Worthwhile noting an example of the oxyl-type ground state for the species in question modelled by simplest neutral complex $FeO(OH)_2$. For this system Malykhin showed that the ferryl-oxyl gap depends strongly on the exchange-correlation potential: for B2PLYP the ground state corresponds to ferryl configuration for the $[FeO]^{2+}$ moiety, while for M06-2X the oxyl state becomes surprisingly lower in energy than ferryl one.[12] Quite unexpectedly the next step, namely, calculation of activation energy for the H-abstraction from methane was performed only for the oxyl excited state with the B2PLYP functional. The obtained barrier of 1.5 kcal/mol appears to be quite small as it should be for a radical-like state. The M06-2X prediction of the oxyl ground state for the $FeO(OH)_2$ complex seems to contradict BP86 predictions of the ferryl ground state for $[FeO]^{2+}$ complexes in zeolite by Baerends and our previous B3LYP results for the same complex.[13],[14] The oxyl ground state for $OFe(OH)_2$ was also obtained by means of the Hartree-Fock-theory based approaches: CCSD(T) [12] and CASSCF [13] which is understandable because the Hartree-Fock exchange generally "prefers" the $d^5$ maximal spin configuration for iron center in the $[FeO]^{2+}$ group.

## Mechanism of the H-abstraction by the $[FeO]^{2+}$ group

Although being formally non-radical in its ground state, the α-oxygen six-membered ring model possesses quite a high reactivity in the H-abstraction from methane, giving a barrier of 6.6 kcal/mol at the DFT/ZORA/BP86/TZ2P level of theory within the ADF package. Worthwhile noting that even a lower barrier of 3 kcal/mol for the same process was found by this group for a charged aqua complex $[FeO(H_2O)_5]^{2+}$.[15] However, for the same complex in the field of uniformly distributed counter charge, this barrier becomes 20 kcal/mol higher.[15] This seems to be a direct evidence for substantial increase of the reactivity via uncompensated charge of the ferryl group.

For all ferryl complexes Baerends with coauthors have been developing the concept that the ferryl-group complexes abstracts hydrogen from hydrocarbons via low-lying alpha-spin $d_{z2}(\sigma^*)$ orbital which is empty in the ferryl configuration $Fe^{IV}=O$. As far as oxyl configuration is concerned, Baerends suggested that in this configuration $d_{z2}(\sigma^*)$ orbital becomes occupied hindering the H abstraction.

Contrary to this point of view the barrier of the H abstraction by the $[FeO]^{2+}$ group in oxyl configuration (for which mentioned $\sigma^*$ orbital is occupied) appears to be substantially lower than that for ferryl configuration as was shown for $[FeO]^{2+}$ group in monomer, dimer and tetramer Fe-hydroxide models [16][17]. This means that despite the fact that oxyl configuration is usually less energetically preferred for the $[FeO]^{2+}$ complexes than ferryl one, in the H-abstraction from methane the energy of the former configurations becomes the ground while approaching the transition state.

Fingerprint of the oxyl state seems to be negligible or negative spin density on terminal oxygen, which accepts the hydrogen from C-H bond. In the course of reaction with hydrogen, the Fe-O bond becomes elongated causing transfer of the α-spin electron from one of doubly occupied π-bonding orbitals leaving behind single β-spin electron in this orbital. This process is schematically plotted in **Ошибка! Источник ссылки не найден.**. The choice of α-spin to be transferred follows the exchange-enhanced reactivity principle formulated by Shaik et al. [18].

For the $[FeO]^{2+}$ in zeolite exchange-cation position S=2 models, Baerends et al found only positive spin density on terminal oxygen in transition state for the H abstraction. Nevertheless, the spin density on "axial" oxygen before the interaction with methane is about twice as much as compared with that for transition state. This means that Baerends' computational data in fact agree with the suggestion on the appearance of oxyl configuration before the actual abstraction of hydrogen.

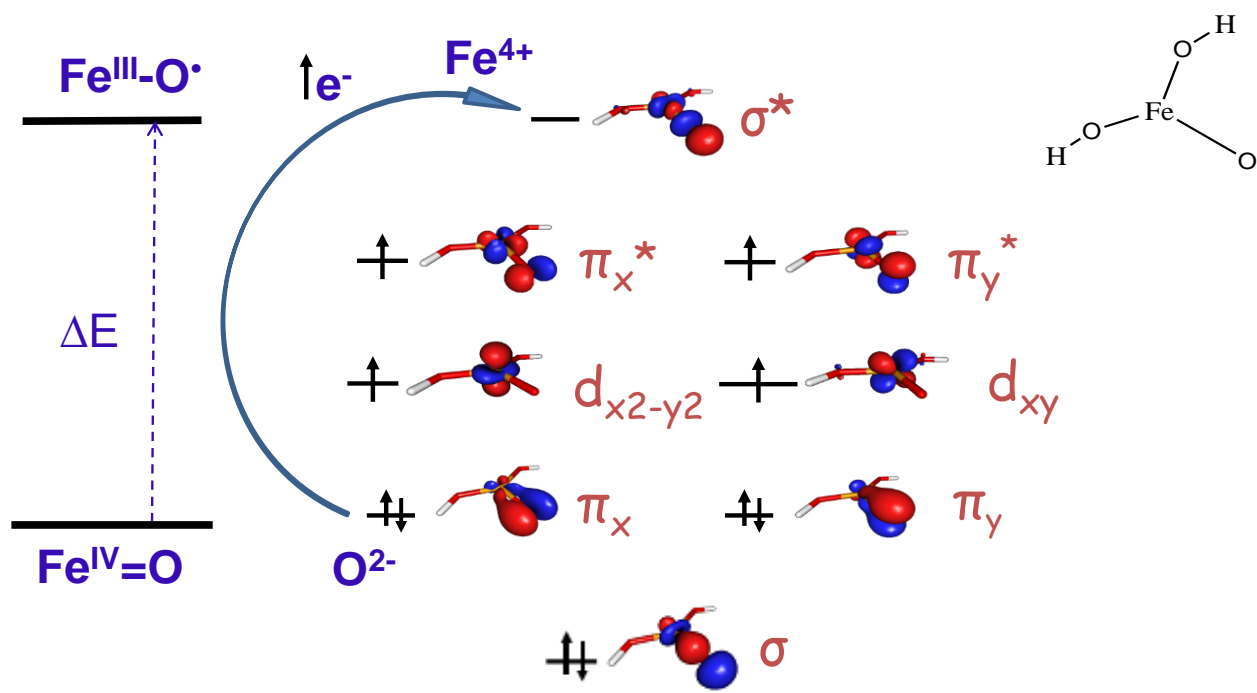

**Figure 1.** Frontier natural orbitals for the [FeO]$^{2+}$ group (from CASSCF(10e,9o) calculation for the simplest FeO(OH)$_2$ model complex [13]) in the ferryl configuration Fe$^{IV}$=O and the oxygen to iron α-spin electron transfer forming the oxyl configuration.

## Ferry-to-oxyl transformation in the reaction course

The [FeO]$^{2+}$ species (coinciding by stoichiometry with the α-oxygen center) was suggested to appear also in the Fe(III)-hydroxides catalyzing the water-to-dioxygen oxidation.[19],[20] These species are assumed to be generated by "external" water oxidizing complex Ru(bpy)$_3^{3+}$ via the abstraction of proton and electron from terminal hydroxo group. In particular, the iron hydroxides γ-FeO(OH) and Fe$_4$(OH)$_{10}$(SO$_4$) appear to be efficient catalysts for the water oxidation.[19] With the use of the di-iron complex Fe$_2$(OH)$_6$ and tetra-iron-hydroxide complex Fe$_4$O$_4$(OH)$_4$ which model the Fe(III) hydroxide, the author of this work with coauthors showed that the O-O bond formation is facilitated in the oxyl type Fe$^{III}$-O$^{\bullet}$ excited state.[21] The same excited-state oxyl group in the mentioned tetramer appears to abstract hydrogen from methane with a barrier as low as 5 kcal/mol while that barrier for the ground-state ferryl group is by a factor of five higher.[16]

In addition to routine computational results, some insight into reactivity of the [FeO]$^{2+}$ group has been obtained via the partition of spin density into the delocalization and polarization contributions in the basis of paired orbitals.[22] A key factor responsible for reactivity of the [Fe-O]$^{2+}$ group was shown to be the spin polarization of terminal oxygen which aligns the approaching hydrogen spin antiparallel to the oxygen spin due to Pauli exclusion principle to form the closed-shell hydroxyl anion. The majority-spin polarization of the ferryl oxygen in $Fe^{IV}=O$ results in majority-spin polarized methyl and the $Fe^{III}$ center in the excited $S = 3/2$ state in the product complex. It is in fact the high lying low-spin state for $Fe^{III}$ that makes the ferryl route unfavorable. The minority-spin polarization of the oxyl oxygen in $Fe^{III}$-O$^•$ forces the methyl group to have the same minority-spin polarization while keeping intact the $S = 5/2$ $Fe^{III}$ state along the reaction pathway. Both described routes are presented schematically in Figure 2.

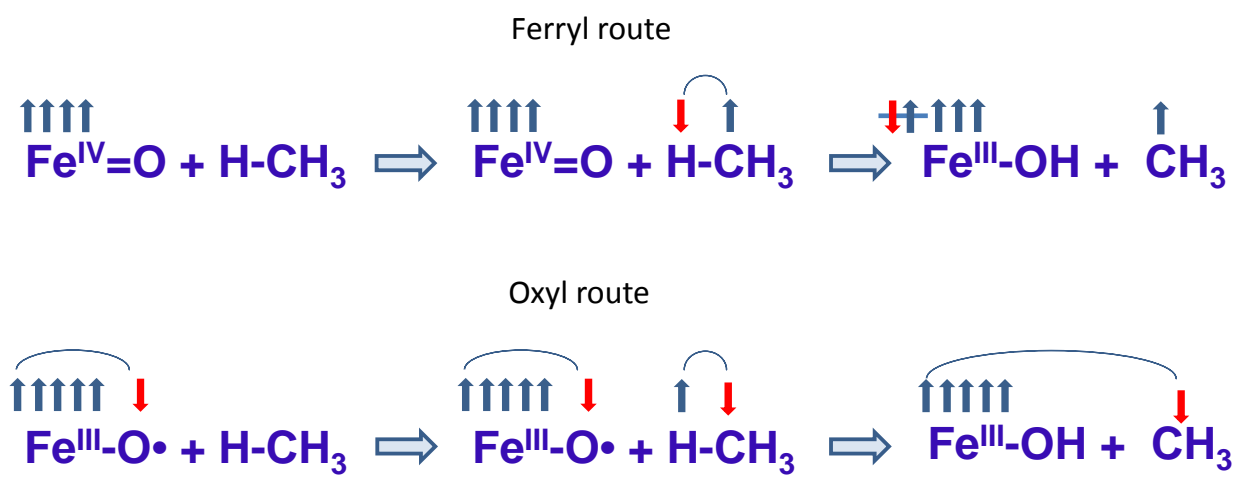

**Figure 2. Simplified scheme of the hydrogen abstraction on the ferryl and oxyl group in the iron hydroxides (modified from** [16] **)**

The electron configuration of the $[FeO]^{2+}$ group in various ligand surrounding is usually of the ferryl type with a negligible contribution of oxyl and only the ligand-to-metal charge transfer excited state possesses the oxyl character. Despite that, in the transition state of the H-abstraction process, the $Fe^{IV}=O$ group transforms to the oxyl $Fe^{III}$-O$^•$ group as was noticed in works of the Solomon's laboratory [23],[24], by Ye and Neese [25], and by Dietl, Schlangen and Schwarz [26],[27] on base of the DFT predictions. In the oxyl transition state, the spin density on reacting oxygen becomes negative as was pointed out by Ye and Neese [25] revealing perhaps the most prominent feature of the oxyl oxygen. Upon the hydroxyl group formation, the β spin shifts from oxyl oxygen to the products, e.g. methyl moiety as was demonstrated for the methane H-abstraction by the $[FeO]^{2+}$ aqueous complex[15], and in our previous work for ferryl containing Fe-hydroxide tetramer complex.[16] Described data imply that the excited oxyl state is a key factor of the $[FeO]^{2+}$ group reactivity. In the course of the reaction, the oxyl term crosses the ferryl term and becomes lower lying state as was shown by Ye and Neese.[25] Moreover, one may suggest that the energy gap between ferryl and oxyl states of the $[FeO]^{2+}$ group determines the barrier of the H abstraction. This mechanism is though effective perhaps only for the ferryl-oxyl gap not larger than some threshold. If the ferryl-oxyl gap is large enough, the proton-coupled electron transfer mechanism takes place which means that only a proton of hydrogen is transferred directly to the accepting oxygen center, while its electron goes to metal via a different route of the same complex (see Schwarz' work[27] and references therein).

# Conclusion

On base of various computational data from literature and those obtained in author's group the mechanism of the H abstraction from methane by the $[FeO]^{2+}$ group can be described as follows. First, the ground state of the $[FeO]^{2+}$ group in the majority of studied complexes is of ferryl $Fe^{IV}$=O type. Second, the H abstraction is preceded by the α-spin electron transfer from oxygen to metal to form oxyl configuration $Fe^{III}$-O$^{\bullet}$. This configuration reveals itself in the rise of spin-polarization component of spin density with negative part localized on oxygen. Third, the methyl fragment appears to bear the β spin which can be considered as a fingerprint of oxyl type route. Finally, Baerends' mechanism is unable to explain the appearance of the β-spin electron on ferryl oxygen in numerous computational data for various $[FeO]^{2+}$ complexes.

[1]  M. Lundberg, T. Borowski, Oxoferryl species in mononuclear non-heme iron enzymes: Biosynthesis, properties and reactivity from a theoretical perspective, Coord. Chem. Rev. 257 (2013) 277–289. doi:10.1016/j.ccr.2012.03.047.

[2]  S.P. de Visser, J.-U. Rohde, Y.-M. Lee, J. Cho, W. Nam, Intrinsic properties and reactivities of mononuclear nonheme iron–oxygen complexes bearing the tetramethylcyclam ligand, Coord. Chem. Rev. 257 (2013) 381–393. doi:10.1016/j.ccr.2012.06.002.

[3]  G.I. Panov, K.A. Dubkov, E. V. Starokon, Active oxygen in selective oxidation catalysis, Catal. Today. 117 (2006) 148–155. doi:10.1016/j.cattod.2006.05.019.

[4]  G.I. Panov, E. V. Starokon, M. V. Parfenov, L. V. Pirutko, Single Turnover Epoxidation of Propylene by α-Complexes (Fe$^{III}$ –O$^{\bullet}$)$_α$ on the Surface of FeZSM-5 Zeolite, ACS Catal. 6 (2016) 3875–3879. doi:10.1021/acscatal.6b00930.

[5]  J.R. Winkler, H.B. Gray, Electronic Structures of Oxo-Metal Ions, in: 2011: pp. 17–28. doi:10.1007/430_2011_55.

[6]  B. Ensing, F. Buda, P.E. Blöchl, E.J. Baerends, A Car–Parrinello study of the formation of oxidizing intermediates from Fenton's reagent in aqueous solution, Phys. Chem. Chem. Phys. 4 (2002) 3619–3627. doi:10.1039/b201864k.

[7]  C. Hammond, M.M. Forde, M.H. Ab Rahim, A. Thetford, Q. He, R.L. Jenkins, et al., Direct catalytic conversion of methane to methanol in an aqueous medium by using copper-promoted Fe-ZSM-5., Angew. Chem. Int. Ed. Engl. 51 (2012) 5129–33. doi:10.1002/anie.201108706.

[8]  G.D. Pirngruber, J.-D. Grunwaldt, P.K. Roy, J.A. van Bokhoven, O. Safonova, P. Glatzel, The nature of the active site in the Fe-ZSM-5/N2O system studied by (resonant) inelastic X-ray scattering, Catal. Today. 126 (2007) 127–134. doi:10.1016/j.cattod.2006.09.021.

[9]  B.E.R. Snyder, P. Vanelderen, M.L. Bols, S.D. Hallaert, L.H. Böttger, L. Ungur, et al., The active site of low-temperature methane hydroxylation in iron-containing zeolites, Nature. 536 (2016) 317–321. doi:10.1038/nature19059.

[10] N.A. Kachurovskaya, G.M. Zhidomirov, E.J.M. Hensen, R.A. van Santen, Cluster Model DFT Study of the Intermediates of Benzene to Phenol Oxidation by N2O on FeZSM-5

Zeolites, Catal. Letters. 86 (2003) 25–31. doi:10.1023/A:1022642521434.

[11] A. Rosa, G. Ricciardi, E.J. Baerends, Is [FeO]2+ the active center also in iron containing zeolites? A density functional theory study of methane hydroxylation catalysis by Fe-ZSM-5 zeolite, Inorg. Chem. 49 (2010) 3866–3880.

[12] S. Malykhin, Hydrogen abstraction reactions of the [FeO]2+ moiety: The role of the electronic state, Chem. Phys. Lett. 622 (2015) 69–74. doi:10.1016/j.cplett.2015.01.024.

[13] I. Zilberberg, R.W. Gora, G.M. Zhidomirov, J. Leszczynski, Bonding in the oxo ferrous iron species: A complete active-space self-consistent-field theory verification of the molecular-oxygen-like pattern, J. Chem. Phys. 117 (2002) 7153. doi:10.1063/1.1506913.

[14] S. Malykhin, I. Zilberberg, G.M. Zhidomirov, Electron structure of oxygen complexes of ferrous ion center, Chem. Phys. Lett. 414 (2005) 434–437.

[15] † Bernd Ensing*, †,‡ Francesco Buda, §. and Michiel C. M. Gribnau, † Evert Jan Baerends*, Methane-to-Methanol Oxidation by the Hydrated Iron(IV) Oxo Species in Aqueous Solution: A Combined DFT and Car−Parrinello Molecular Dynamics Study, (2004).

[16] A.A. Shubin, S.P. Ruzankin, I.L. Zilberberg, V.N. Parmon, Distinct activity of the oxyl FeIIIO group in the methane dissociation by activated iron hydroxide: DFT predictions, Chem. Phys. Lett. 640 (2015) 94–100. doi:10.1016/j.cplett.2015.10.016.

[17] V. Kovalskii, A.A. Shubin, Y. Chen, D. Ovchinnikov, S.P. Ruzankin, J. Hasegawa, et al., Hidden radical reactivity of the [FeO]2+ group of the (hydro)oxide species in the H-abstraction from methane: a DFT and CASPT2 study, Chem. Phys. Lett. 679 (2017) 193–199. doi:10.1016/j.cplett.2017.05.002.

[18] S. Shaik, H. Chen, D. Janardanan, Exchange-enhanced reactivity in bond activation by metal–oxo enzymes and synthetic reagents, Nat. Chem. 3 (2011) 19–27. doi:10.1038/nchem.943.

[19] G.L. Elizarova, G.M. Zhidomirov, V.N. Parmon, Hydroxides of transition metals as artificial catalysts for oxidation of water to dioxygen, Catal. Today. 58 (2000) 71–88. doi:10.1016/S0920-5861(00)00243-1.

[20] M.J. Filatov, G.L. Elizarova, O.V. Gerasimov, G.M. Zhidomirov, V.N. Parmon, Quantum-chemical study of the four-electron catalytic oxidation of water to dioxygen in the presence of dinuclear and tetranuclear hydroxide complexes of cobalt(III) and iron(III): Intermediates of the catalytic cycle and their relative energies, J. Mol. Catal. 91 (1994) 71–82. doi:10.1016/0304-5102(94)00014-X.

[21] A.A. Shubin, S.P. Ruzankin, I.L. Zilberberg, O.P. Taran, V.N. Parmon, The routes of association of (hydro)oxo centers on iron hydroxide at the water oxidation process: DFT predictions, Chem. Phys. Lett. 619 (2015) 126–132. doi:10.1016/j.cplett.2014.11.061.

[22] S.P. Ruzankin, I. Lyskov, I.L. Zilberberg, Net spin and polarization components of the spin density for the single determinant in the basis of paired orbitals, Int. J. Quantum Chem. 112 (2012) 3052–3058.

[23] M.L. Neidig, A. Decker, O.W. Choroba, F. Huang, M. Kavana, G.R. Moran, et al., Spectroscopic and electronic structure studies of aromatic electrophilic attack and hydrogen-atom abstraction by non-heme iron enzymes, Proc. Natl. Acad. Sci. 103 (2006) 12966–12973. doi:10.1073/pnas.0605067103.

[24] M. Srnec, S.D. Wong, J. England, L. Que, E.I. Solomon, π-Frontier molecular orbitals in S = 2 ferryl species and elucidation of their contributions to reactivity., Proc. Natl. Acad. Sci. U. S. A. 109 (2012) 14326–31. doi:10.1073/pnas.1212693109.

[25] S. Ye, F. Neese, Quantum chemical studies of C-H activation reactions by high-valent nonheme iron centers, Curr. Opin. Chem. Biol. 13 (2009) 89–98. doi:10.1016/j.cbpa.2009.02.007.

[26] N. Dietl, M. Schlangen, H. Schwarz, Thermal hydrogen-atom transfer from methane: the role of radicals and spin states in oxo-cluster chemistry., Angew. Chem. Int. Ed. Engl. 51 (2012) 5544–55. doi:10.1002/anie.201108363.

[27]  H. Schwarz, Thermal hydrogen-atom transfer from methane: A mechanistic exercise, Chem. Phys. Lett. 629 (2015) 91–101. doi:10.1016/j.cplett.2015.04.022.